\documentclass[prb, twocolumn, nofootinbib]{revtex4-2}
\usepackage{amssymb}
\usepackage{amsmath}
\usepackage{graphicx}
\usepackage{bm}
\usepackage{xcolor}
\usepackage[unicode=true,colorlinks=true]{hyperref}
\hypersetup{  
     urlcolor = blue }

\begin{document}

\title{Tight-binding description of inorganic lead halide perovskites in cubic phase}

\author{M.~O.~Nestoklon}
\affiliation{Ioffe Institute, St.~Petersburg 194021, Russia}

\begin{abstract}

Band structure of inorganic lead halide perovskites is substantially different from the band structure of group IV, III-V and II-VI semiconductors. However, the standard empirical tight-binding model with $sp^3d^5s^*$ basis gives nearly perfect fit of band structure calculated in the density functional theory.  The tight-binding calculations of ultrathin CsPbI\textsubscript{3} layer show good agreement with the corresponding DFT calculations. The parameters corrected for the experimental data allow for the numerically cheap atomistic calculations of inorganic perovskite nanostructures.

\end{abstract}


\maketitle

\section{Introduction}

Lead halide perovskites attracted a lot of attention last decade due to excellent optical properties and simple processing \cite{Xing14,Green14,Protesescu15,Yakunin15}. Clearly, design of new perovskite-based structures and devices calls for theoretical models which would adequately describe their properties. While modern methods of electron structure calculations allow one to compute the band structure of bulk semiconductors with reasonable precision \cite{Heyd06}, the use of density functional theory (DFT) for nanostructure calculations is challenging, and more computationally accessible approaches are required. For large structures and qualitative analysis, effective mass approximation \cite{Luttinger55} is the method of choice. For intermediate size nanostructures,  the attraction is toward atomistic empirical methods \cite{Zunger_PP,Xavier_book}. Empirical tight-binding method (ETB) is one of the simplest approximations suitable for an accurate description of the band structure of conventional semiconductors with lowest possible computational cost \cite{Xavier_book}. 

The band structure of inorganic perovskites and hybrid organic-inorganic perovskites is substantially different from the band structure of widely used group IV, III--V, and II--VI semiconductors. First of all, bulk perovskites have a complex sequence of structural phase transitions. To make the analysis simpler, it is natural to first consider the high-symmetry cubic structure and then describe energy bands of lower symmetry crystal phases as originating from a folded and distorted band structure of the most symmetric phase \cite{Even15}. The band structure of the cubuc phase is also important by itself: recently it has been demonstrated that, in small nanocrystals, the cubic phase is stabilized by the surface up to room temperature \cite{Liu19}. In cubic perovskites, the direct band gap is located at the R point of the Brillouin zone which has $O_h$ point symmetry. The top of the valence band is formed by the states which transform under $\Gamma_6^-$ representation (in Koster notation \cite{bookKoster}), while the bottom of the conduction band is formed by $\Gamma_6^+$ states \cite{Soline16}. The difference with III--V and II--VI direct band gap semiconductors is in the formation of these bands: in III--V and II-VI compounds, the spin-orbit interaction forms the valence band states and is negligible in the conduction band while, in perovskites, the conduction band is formed by the large spin-orbit interaction. In perovskites, both the valence band top and conduction band bottom are two-fold degenerate while the top of the valence band in III-V and II-VI compounds is four-fold degenerate which results in much simpler selection rules for the optical transitions. 

The empirical tight-binding method with $sp^3d^5s^*$ basis in the nearest neighbor approximation gives a precise description of the band structure of bulk III--V \cite{Jancu98} and group IV \cite{Niquet09} semiconductors. It has a long history of application to nanostructures \cite{Robert12,Poddubny12,Nestoklon16_Ge,Robert16,Benchamekh16,Tan16}. Parameters of the method may be in principle extracted from the DFT calculations \cite{Tan15,Gresch18,Lihm19} which makes this method an atomistic interpolation of the underlying DFT model. It may be extended to describe other materials as well but, for perovskites, ETB has been used mostly for a qualitative analysis of the band structure \cite{Soline16,DiVito20}. Below we show that it can be used to describe the band structure of inorganic perovskites with a meV-range precision. 

The empirical methods of band structure calculations heavily rely on the band structure they are assumed to fit. For this purpose we use the band structure resulting from DFT calculations with the modified Becke-Johnson exchange-correlation potential \cite{Tran09} in Jishi parametrization \cite{Jishi14}. While the band-gap energy is known to be underestimated in most DFT calculations, this exchange-correlation potential is almost free from this deficiency \cite{Camargo12} while it may give a noticeable error in masses and positions of secondary maxima in the conduction band \cite{Wang13}. However, the standard DFT calculations give the ``bare'' electron structure without account on the electron-phonon interaction. In many semiconductors, the renormalization of the band structure by the electron-phonon interaction is within few tens meV \cite{Cardona01}, which is beyond standard precision of DFT calculations. However, for lead halide perovskites this effect is estimated to be of the order of hundreds meV \cite{Wiktor17}. This value is also large in other perovskites \cite{Patrick15,Zacharias20}. To account for this systematic error, we use the approach which  used to be the ``golden standard'' when high-quality electron structure calculations were unavailable. First, we obtain the best possible fit of the band structure obtained in DFT calculations. Next, we consider a thin perovskite slab and compare the in-plane dispersion of the electron states in this structure calculated in DFT and ETB. Nearly perfect agreement between the two approaches shows that ETB may be used for the calculations of the perovskite nanostructures. Finally, we revise this parametrization in order to introduce corrections aimed to describe the experimental data on the material band structure. This procedure gives the two sets of parameters: one is suitable for a direct comparison with the DFT calculations, the other one may be used in nanostructure calculations to be compared with the experimental data. 

\section{Density functional calculations of cubic \NoCaseChange{CsPbI}\texorpdfstring{\textsubscript{3}}{3}}\label{sec:DFT}


For the band structure calculations we use the WIEN2k package \cite{wien2k}. For better comparison with experimental data, we use the modified Becke-Johnson exchange-correlation potential \cite{Tran09} in Jishi parametrization \cite{Jishi14}. The lattice constant of cubic CsPbI\textsubscript{3} is set to $a=6.289$~\AA. All parameters of the convergence are default except $R_{MT}\cdot K_{max}=9$, $l_{vns}=6$, and the FFT mesh enhancement factor 4. In addition, for spin-orbit interaction RLO \cite{Kunes01} is added on Pb atoms and \texttt{emax}=10 is used. We use $k$-mesh $14\times 14\times 14$. The energy and charge convergence were set to $10^{-4}$ and $10^{-3}$, respectively. With these convergence parameters, for the band gap we obtain the value $E_g=1.017$~eV. Note that the experimental value of the band gap of CsPbI\textsubscript{3} in the cubic phase is $E_g=1.65$~eV \cite{Yuan20}. The large difference between DFT results and experimental data for this material is usually attributed to the renormalization of the band structure by the electron-phonon interaction \cite{Cardona01}. For lead halide perovskites this effect is estimated to be of the order of hundreds meV \cite{Wiktor17}. 

In WIEN2k, Kohn-Sham equations are solved in the basis which is obtained as a numerical combination of Slater-like numerical orbitals near atoms and plane waves between the spheres around atoms. For the valence band states, most of electron density is located near the atoms which allows for an unambiguous definition of the angular momentum-resolved partial density of states. This is in line with the general idea of the tight-binding method which assumes that the band structure may be obtained in the basis of (in principle, unknown) functions localized near atoms and corresponding to a certain value of the orbital angular momentum \cite{Slater54}. From the analysis of DFT calculations, one may conclude that the valence band is mostly constructed from functions with orbital angular momentum 1 (``$p$-orbitals'' in ETB) on halogen (I in case of CsPbI\textsubscript{3}) atoms with significant admixture of functions with zero momentum (``$s$-orbitals'') on Pb. Conduction band is constructed almost exclusively from functions with orbital angular momentum 1 on Pb. Cs atoms give few flat bands deep inside valence band and we drop them from the ETB parametrization, following Ref.~\cite{Soline16}. This analysis gives a good starting point for the ETB parametrization without explicit interpolation between DFT and ETB wave functions \cite{Gresch18}.

\section{Interpolation of DFT band structure of cubic \NoCaseChange{CsPbI}\texorpdfstring{\textsubscript{3}}{3} in empirical tight-binding}

\subsection{ETB parametrization of DFT results in \texorpdfstring{$sp^3$}{sp3} basis}

The use of complete $sp^3d^5s^*$ model is critically important to obtain the correct energy positions of the bands in all high-symmetry points of the Brillouin zone, see below. However, it is instructive to start from the minimal model and check if the band structure may be fit to be in  agreement with the DFT results. To fit the ETB parameters, we begin with a straightforward minimization of the difference between DFT and ETB energies in special points of the Brillouin zone starting from the parameters of Ref.~\cite{Soline16}. The starting parameters are important to have the correct order of the bands. In this case the procedure works even without special attention to local minima. After this general fit, which indeed gives band structure similar to the DFT results, we refine the ETB parameters to exactly reproduce few selected energies in the special points calculated in the DFT. First we fit the top of the valence band in $R$ point, then the bottom of the conduction band in $R$ point, then position of Pb $s$-band in $R$ point (band near $-8$~eV), then position of the lowest I $p$-band in $\Gamma$ point (band near $-4$~eV), and finally the position of the conduction band bottom in $M$ point. The result of the fit is presented in Fig.~\ref{fig:TBsp3vsDFT}. The spin-orbit interaction is taken from the $sp^3d^5s^*$ fit, see below.

\begin{figure}
\includegraphics[width=\linewidth]{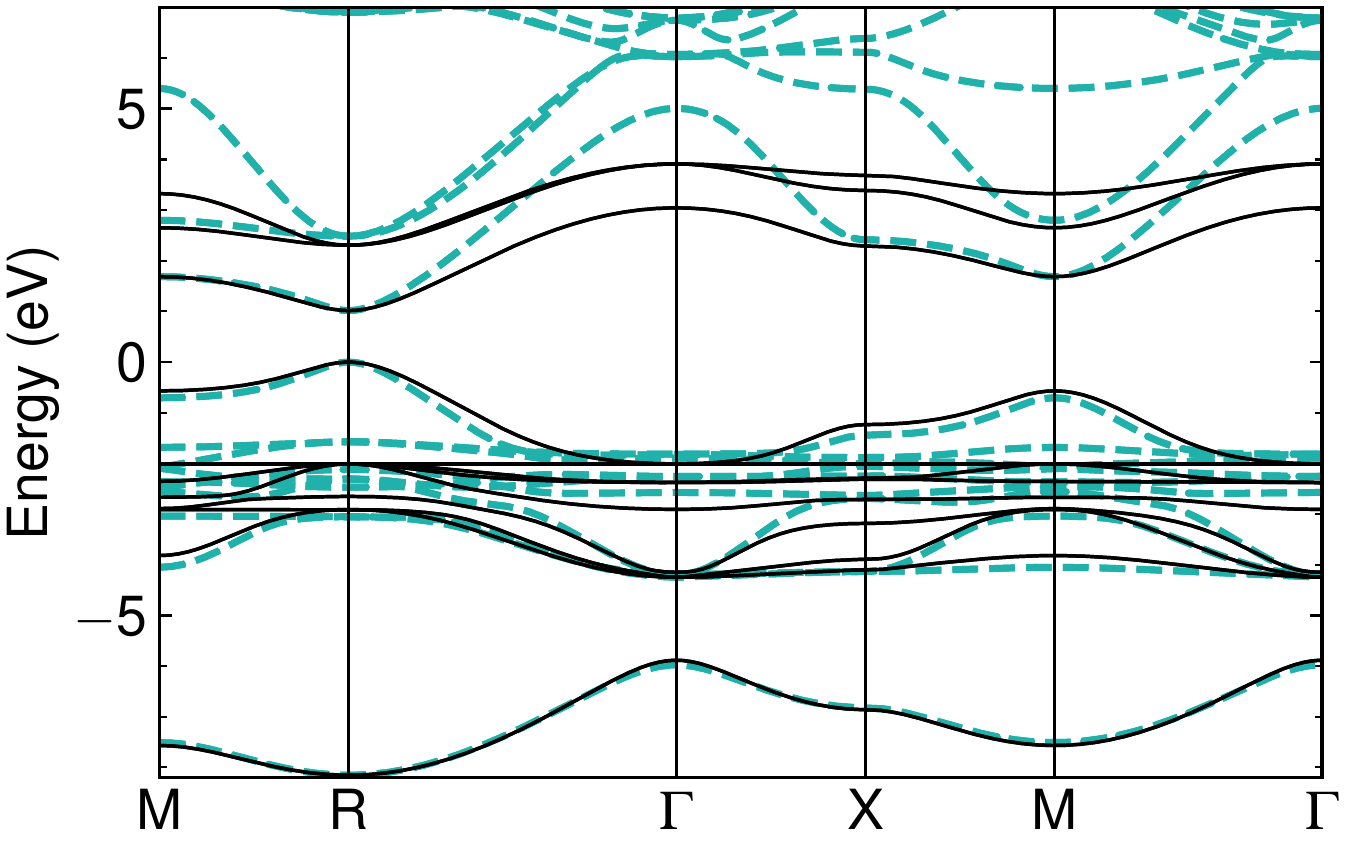}
 \caption{
   Comparison of CsPbI\textsubscript{3} band structure calculated in DFT and in empirical tight-binding method using $sp^3$ model with parameters from Table~\ref{tbl:TB_par}. Green dashed lines show the DFT results, black solid lines ETB results.
}\label{fig:TBsp3vsDFT}
\end{figure}

Within the minimal $sp^3$ model, it is impossible to further increase the number of points to be fitted exactly. In particular, note the incorrect position of the bottom of the conduction band in $\Gamma$ point which arises from the absence of higher bands in the minimal ETB model. However, the overall description of the band structure is surprisingly good. In particular, this model ``automatically'' reproduces the position of the valence band maximum in $M$ point and the splitting of bands in the region from $-4$ to $-3$~eV in the valence band. The structure of the bands in the region from $-3$ to $-2$~eV is not reproduced exactly.

 \begin{table}\caption{Tight-binding parameters used in calculations. In addition to parameters presented in the table, $s^*s^*\sigma=s_cs^*_a\sigma=s^*_ad_c\sigma=p_cd_a\sigma=p_ad_c\pi=p_cd_a\pi=0$.}
\label{tbl:TB_par}
 \begin{tabular*}{\linewidth}{@{\extracolsep{\fill}}lrrr}
 \hline
 \hline
 &$sp^3$ & $sp^3d^5s^*$ & expt. corrected \\
 \hline
$         E_{sa}$ &$ -13.387  $   & $  -13.3903$ & $  -13.3903$\\
$         E_{sc}$ &$  -5.885  $   & $   -5.8406$ & $   -5.8406$\\
$       E_{s^*a}$ &$          $   & $   19.8877$ & $   19.7828$\\
$       E_{s^*c}$ &$          $   & $   22.0026$ & $   22.0026$\\
$         E_{pa}$ &$  -2.310  $   & $   -2.2838$ & $   -2.2838$\\
$         E_{pc}$ &$   1.774  $   & $    3.9373$ & $    4.6044$\\
$         E_{da}$ &$          $   & $   10.9883$ & $   10.9883$\\
$         E_{dc}$ &$          $   & $   13.9810$ & $   13.9810$\\
$       ss\sigma$ &$  0.003   $   & $   -0.2570$ & $   -0.2570$\\
$ s_as^*_c\sigma$ &$          $   & $    0.5198$ & $    0.5198$\\
$   s_ap_c\sigma$ &$  0.627   $   & $    0.0047$ & $    0.0155$\\
$   s_cp_a\sigma$ &$  1.048   $   & $    1.0565$ & $    1.0565$\\
$ s^*_ap_c\sigma$ &$          $   & $    2.9449$ & $    2.9131$\\
$ s^*_cp_a\sigma$ &$          $   & $   -0.3643$ & $   -0.3642$\\
$   s_ad_c\sigma$ &$          $   & $    0.0290$ & $    0.0290$\\
$   s_cd_a\sigma$ &$          $   & $    0.5825$ & $    0.5825$\\
$ s^*_cd_a\sigma$ &$          $   & $    0.0859$ & $    0.0859$\\
$       pp\sigma$ &$ -1.376   $   & $   -1.7915$ & $   -1.8988$\\
$          pp\pi$ &$  0.635   $   & $    0.5421$ & $   -0.0331$\\
$   p_ad_c\sigma$ &$          $   & $    1.0063$ & $    1.0063$\\
$       dd\sigma$ &$          $   & $   -1.0285$ & $   -1.0285$\\
$          dd\pi$ &$          $   & $    2.0000$ & $    2.0000$\\
$       dd\delta$ &$          $   & $   -1.4000$ & $   -1.4000$\\
$     \Delta_a/3$ &$  0.300   $   & $    0.2836$ & $    0.2836$\\
$     \Delta_c/3$ &$  0.433   $   & $    0.5413$ & $    0.5413$\\
 \hline
 \hline
 \end{tabular*}
 \end{table}

In the next section we use the $sp^3d^5s^*$ model, but a brief comment should be added concerning the use of an intermediate $sp^3s^*$ model. Similar to simple direct band gap  semiconductors \cite{Vogl83}, the $sp^3s^*$ model allows one to fit the position of the conduction band bottom in $\Gamma$ point. However, the behavior of the second conduction band is not reproduced even qualitatively and we skip a detailed discussion of this model. 

\subsection{ETB parametrization of DFT results in \texorpdfstring{$sp^3d^5s^*$}{sp3d5s} basis}\label{sec:TB_sp3d5s}

\begin{figure}
\includegraphics[width=\linewidth]{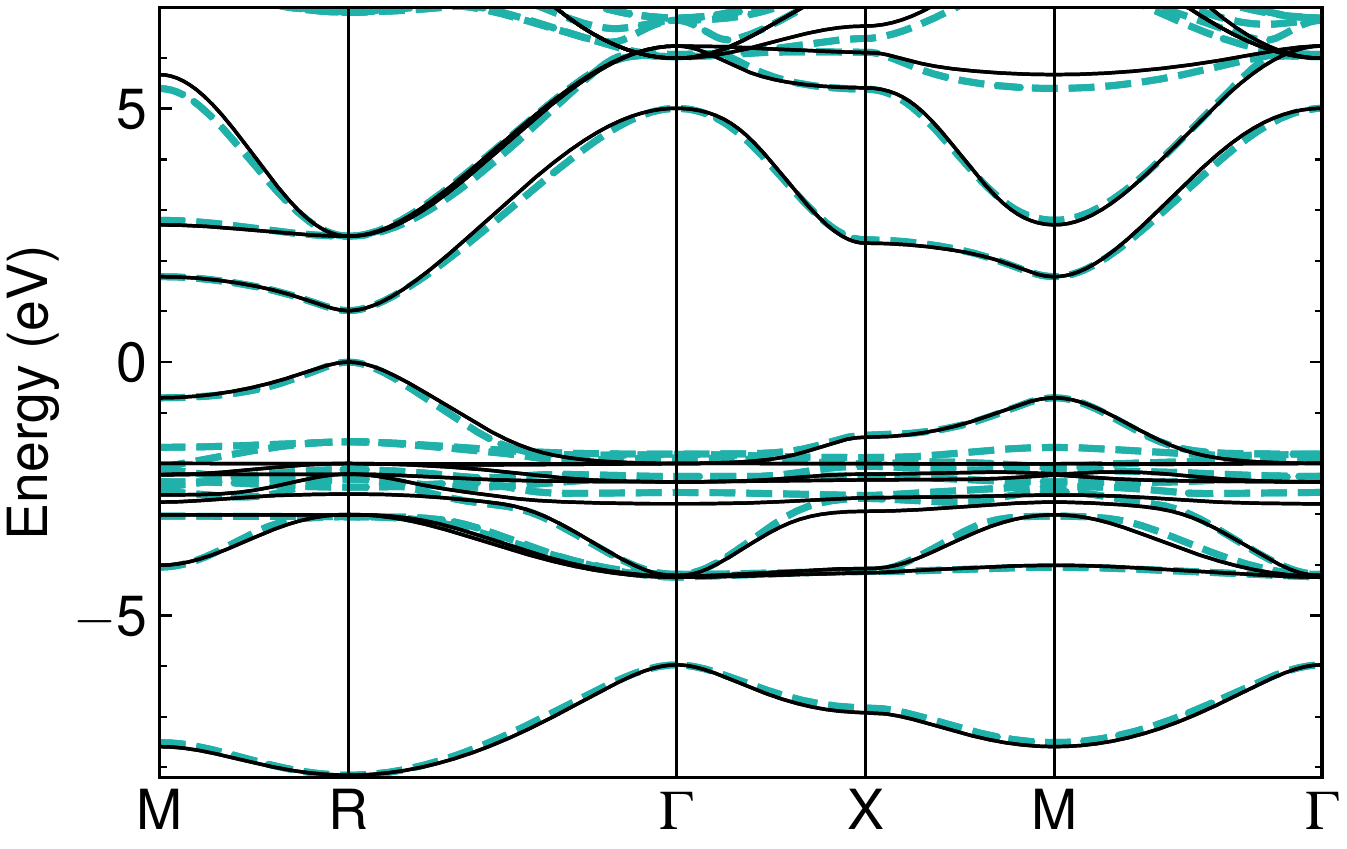}
 \caption{
   Comparison of CsPbI\textsubscript{3} band structure calculated in DFT and in empirical tight-binding method. The DFT calculations are shown in green dashed lines, tight-binding results are in thin black lines. 
}\label{fig:TBvsDFT}
\end{figure}

The minimal $sp^3$ model does not reproduce the band structure of the bands near the band gap far from the $R$ point. This is of critical importance for an accurate description of the nanostructures as the details of the band structure in the whole Brillouin zone are important \cite{Belolipetsky18}. We follow the strategy proved to be efficient for III-V semiconductors \cite{Jancu98}: we start from the $sp^3$ parameters fitted to the DFT band structure and add $s^*$ and $d$ orbitals in the model with energies close to the energies expected from the free-electron dispersion. Then, we reoprimize the parameters allowing for the interaction between the $sp^3$ and $d^5s^*$ orbitals. In this case, general optimization of the parameters is not necessary and we may start directly from the exact fit of few selected special points. The result of this improved parametrization is shown in Fig.~\ref{fig:TBvsDFT}. The resulting fit is nearly perfect, with the exception of the bands in the region from $-3$ to $-2$~eV. It may be shown that inclusion of the interaction with distant bands is not enough to properly split these bands and an exact fit is impossible within the nearest neighbor $sp^sd^5s^*$ model. To show this, one may numerically evaluate the Jacobian matrices $\frac{\partial E_i(k_k)}{\partial p_j}$ of the derivatives of energies in the special points of the Brillouin zone with respect to the ETB parameters and see that there are no parameters which would allow one to split these bands in the correct order. It may be shown that this is not the local optimum by considering small, but not negligible change of parameters. For the exact ETB description of these bands the use of second neighbor interaction or a generalization of the Slater-Koster scheme \cite{Slater54} is necessary.

\section{Tight-binding description of surface of cubic \NoCaseChange{CsPbI}\texorpdfstring{\textsubscript{3}}{3}}

The main aim of the current study is the ETB parameters which are suitable for the description of nanostructures from inorganic halide perovskites. In the previous section it has been demonstrated that the description of the band structure of bulk CsPbI\textsubscript{3} with nearest neighbor $sp^3d^5s^*$ ETB is possible. For the nanostructures, the description of the surface is also important. To check that the description of the surface is reasonable, one should compare the tight-binding results with DFT calculations for a structure with a surface. 

\begin{figure}
\hfil\raisebox{-0.5\height}{\includegraphics[width=0.66\linewidth]{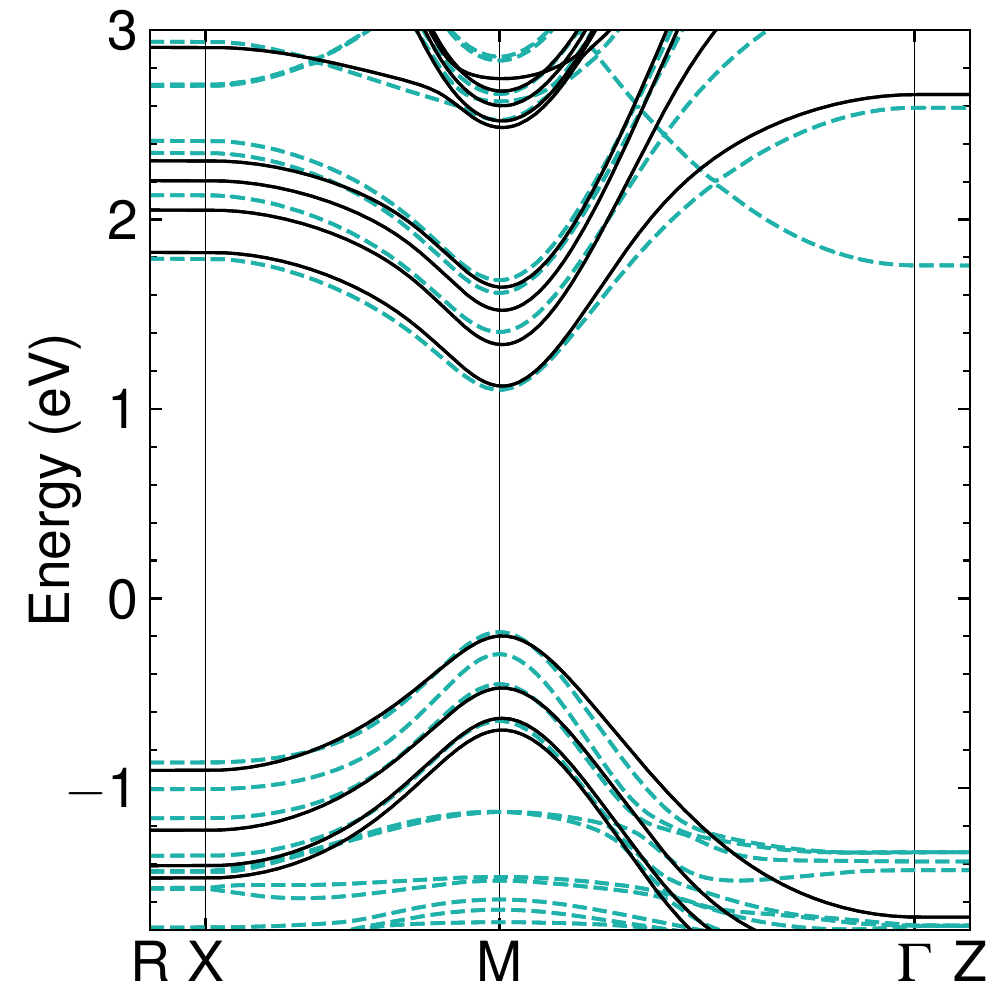}}\hfil%
\raisebox{-0.5\height}{\includegraphics[width=0.25\linewidth]{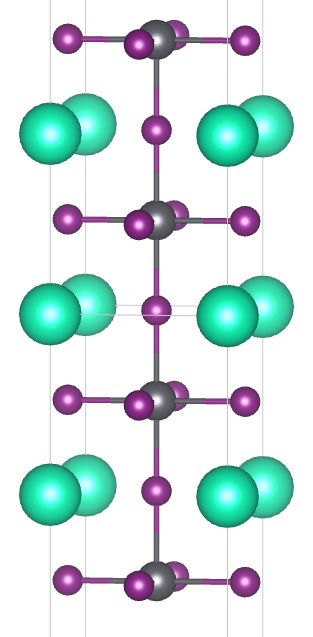}}\hfil%
 \caption{
   Band dispersion in CsPbI\textsubscript{3} nanoplatelet calculated in DFT following the procedure outlined in Sec.~\ref{sec:DFT} is shown in green dashed line and the dispersion calculated in ETB with the parameters obtained in Sec.~\ref{sec:TB_sp3d5s} is shown in black solid lines. To the right, 3D view \cite{momma2011vesta} of arrangement of atoms in elementary cell of CsPbI\textsubscript{3} platelet used for the comparison of DFT and ETB.
}\label{fig:DFT_surface}
\end{figure}

For this purpose, we calculate the dispersion in a CsPbI\textsubscript{3} nanoplatelet in DFT using the same approach and convergence parameters we used for the band structure calculations of bulk CsPbI\textsubscript{3} in Section~\ref{sec:DFT}. An elementary cell of the calculated structure is shown in Fig.~\ref{fig:DFT_surface}. To model the surface we add 14~\AA\ of vacuum in the ``$z$'' direction. We set constant parameters $c$ \cite{Tran09} taken from the converged bulk band structure calculation.
 
Properties of the halide perovskite surface are sensitive to details of the surface passivation \cite{Yang19}. Without passivation, a structure with an iodine surface in DFT is metallic. However, a structure with a surface abruptly cut at the PbI plane demonstrates the band structure one would expect from a quantum-well like structure with quantum confined states localized in the middle of the platelet. The band structure of such a system calculated in DFT is shown in Fig.~\ref{fig:DFT_surface}. To simplify comparison with the ETB results we neglect the structure relaxation. In DFT calculations with structure relaxation (not presented here), the surface atoms are noticeably shifted, but the band structure remains qualitatively the same, with some band gap change. To compare the DFT results for the relaxed structure with ETB calculations we need to incorporate the deformation in TB model, which is beyond the scope of the current paper. 

The band structure of this slab calculated using ETB with the parameters obtained in the previous Section are also presented in Fig.~\ref{fig:DFT_surface}. As seen from the comparison of the two calculations, ETB qualitatively correctly describes the surface. For a detailed model of a realistic system with a particular passivation the procedure presented here should be significantly extended allowing for the passivation within the ETB model. For any particular passivation one might need a separate set of the surface ETB parameters. However, this would heavily rely on the DFT calculations of the passivated surface which is complicated by the fact that most of the standard passivation agents used in experiments (oleic acid, etc. \cite{Yang19}) are relatively large organic molecules. We also neglect the effect of dielectric confinement \cite{Katan19}. Note, however, that the effect of dielectric confinement is usually canceled out by the excitonic effects \cite{Benchamekh14,Cho19}.

\section{ETB parameters corrected for the experimental data}

In previous sections, it has been demonstrated that ETB gives nearly perfect fit of cubic CsPbI\textsubscript{3} band structure calculated in DFT. In addition, even without special parametrization, it gives a reasonable description of the electronic structure of the surface of this material. However, the agreement between resulting ETB and experimental data is not satisfactory, due to the large difference between the underlying DFT results and experiment. To be able to compare ETB results with experimental data, we alter the ETB parameters obtained from the fit of the DFT band structure to reproduce experimental data on the energy bands of cubic CsPbI\textsubscript{3}. 

\begin{figure}[tb]
\includegraphics[width=\linewidth]{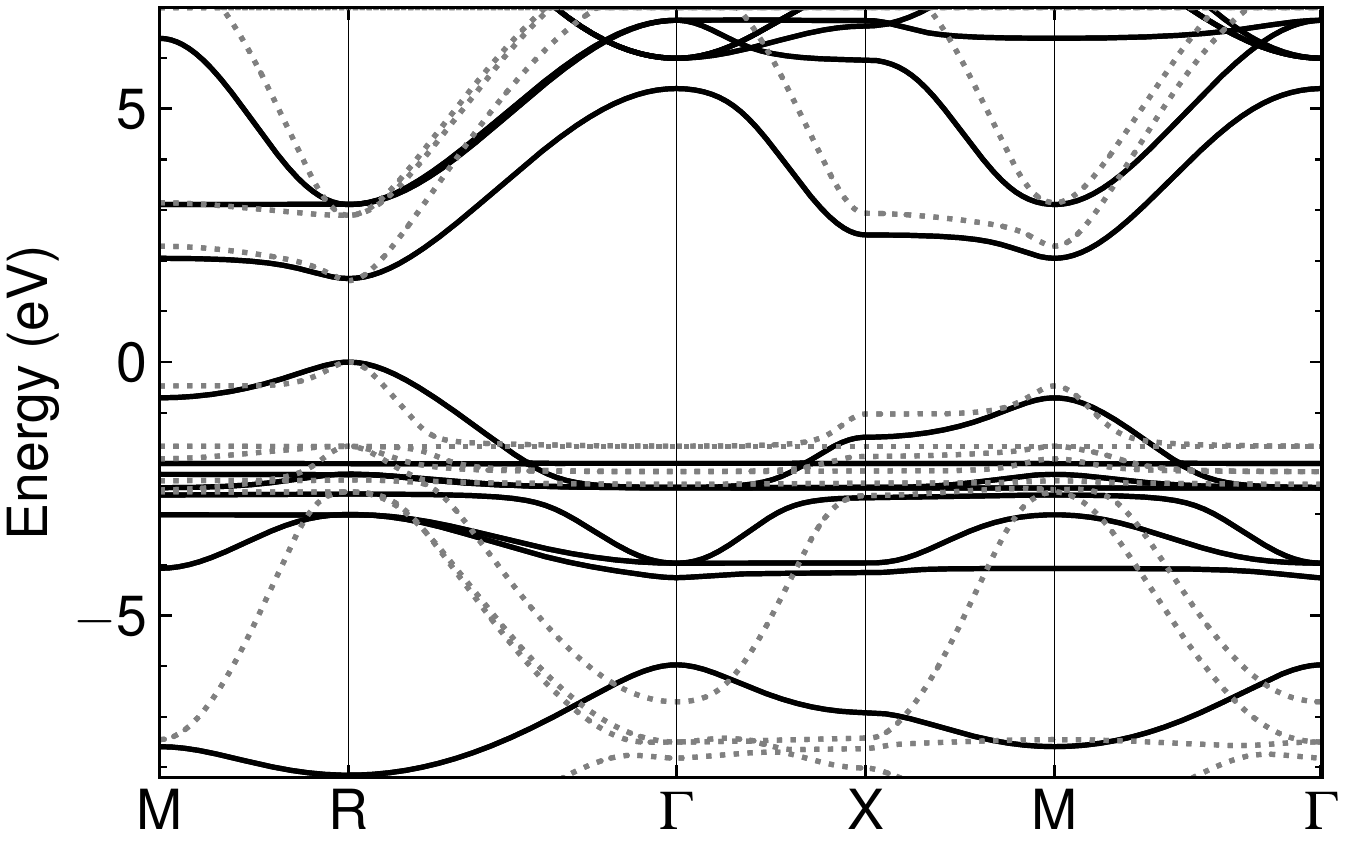}
 \caption{
   The band structure calculated using the ETB parameters fit to DFT results and corrected for the experimental data. Grey dotted line shows the band structure calculated using parameters from Ref.~\cite{Soline16}.
}\label{fig:TBvsExpt}
\end{figure}

For the experimental data we refer to Ref.~\cite{Yuan20}, in this work authors associate the three peaks in optical absorption of $\alpha$-phase (cubic) CsPbI\textsubscript{3} at 1.65~eV, 2.75~eV and 3.40~eV with the direct transitions between the valence and conduction band in R point, in M point and in X point of the Brillouin zone, respectively. In our DFT calculations, the spin-orbit splitting of the conduction band in R point is 1.48~eV. The value of spin-orbit splitting is rather stable against any perturbations including a band gap change, which means that the direct transition in R point should be accompanied by a transition which is 1.48~eV above. It allows us to assume that the third peak observed in the experiment \cite{Yuan20} corresponds to the direct transition in R point from the top of the valence band to the second conduction band. Then, the energies of all the direct transitions in our DFT calculations are underestimated for $400\div600$~meV as compared with the experimental data which we associate with neglect of the electron-phonon interaction \cite{Cardona01}. Based on this consideration, we change the target position of the conduction band bottom in R and M to have the direct band gaps in these points equal to $E_g^{R} = 1.65$~eV and $E_g^{M} = 2.75$~eV. We set the new values of energies in special points of the Brillouin zone and change the parameters accordingly using the procedure outlined above. The results are presented in Fig.~\ref{fig:TBvsExpt}. For comparison, we included also ETB calculations using parameters from Ref.~\cite{Soline16}.

It should be noted, that the experimental data in Ref.~\cite{Yuan20} were obtained at relatively high temperature: bulk CsPbI\textsubscript{3} undergoes phase transitions to lower symmetry phases at 260$\phantom{|}^{\circ}$C and 160$\phantom{|}^{\circ}$C. Our main goal is to find the parameters suitable for the ETB calculations of perovskite quantum dots at room temperature. It has been demonstrated experimentally that, in small QDs, a (meta)stable cubic phase of CsPbI\textsubscript{3} may exist down to room temperature and below \cite{Liu19}. This stabilization is associated with the balance between the bulk and surface energies of different phases \cite{Liu19,YangWang19}. We hypothesize that the mechanism of the cubic phase stabilization should have an effect on CsPbI\textsubscript{3} band gap similar to that of the elevated temperature. Under this assumption, we use the experimental band gap of the cubic perovskite at high ($\sim$300$\phantom{|}^{\circ}$C) temperature to fit the ETB band structure corrected for the experiment. Also note that in these materials the change of the band gap with the temperature is surprisingly small as compared with the zero-temperature correction.




\section{Conclusions}
In conclusion, the band structure of the inorganic cubic perovskite CsPbI\textsubscript{3} calculated using the density functional approach has been fitted near the band gap with a meV precision by the standard nearest-neighbor $sp^3d^5s^*$ empirical tight-binding model. Description of complex crystal-field splitting of few bands formed from the $p$-orbitals of halogen atoms is problematic for the nearest-neighbor model and an extension of the model is necessary for a precise description of these flat valence bands. 

It has been demonstrated that the empirical tight-binding gives a qualitatively correct description of the PbI [001] surface of CsPbI\textsubscript{3} without passivation, which opens the way to numerically cheap simulations of inorganic perovskite based nanostructures. A set of empirical tight-binding parameters carefully fit to the available experimental data has been obtained. 

\section*{Acknowledgments}
Author thanks M.M. Glazov and S.V. Goupalov for helpful discussions. 
The work was supported by RFBR grant 19-02-00545 and the Foundation for advancement of theoretical physics and mathematics ``BASIS''.


\bibliography{perovskites_TB}

\begin{thebibliography}{45}%
\makeatletter
\providecommand \@ifxundefined [1]{%
 \@ifx{#1\undefined}
}%
\providecommand \@ifnum [1]{%
 \ifnum #1\expandafter \@firstoftwo
 \else \expandafter \@secondoftwo
 \fi
}%
\providecommand \@ifx [1]{%
 \ifx #1\expandafter \@firstoftwo
 \else \expandafter \@secondoftwo
 \fi
}%
\providecommand \natexlab [1]{#1}%
\providecommand \enquote  [1]{``#1''}%
\providecommand \bibnamefont  [1]{#1}%
\providecommand \bibfnamefont [1]{#1}%
\providecommand \citenamefont [1]{#1}%
\providecommand \href@noop [0]{\@secondoftwo}%
\providecommand \href [0]{\begingroup \@sanitize@url \@href}%
\providecommand \@href[1]{\@@startlink{#1}\@@href}%
\providecommand \@@href[1]{\endgroup#1\@@endlink}%
\providecommand \@sanitize@url [0]{\catcode `\\12\catcode `\$12\catcode
  `\&12\catcode `\#12\catcode `\^12\catcode `\_12\catcode `\%12\relax}%
\providecommand \@@startlink[1]{}%
\providecommand \@@endlink[0]{}%
\providecommand \url  [0]{\begingroup\@sanitize@url \@url }%
\providecommand \@url [1]{\endgroup\@href {#1}{\urlprefix }}%
\providecommand \urlprefix  [0]{URL }%
\providecommand \Eprint [0]{\href }%
\providecommand \doibase [0]{https://doi.org/}%
\providecommand \selectlanguage [0]{\@gobble}%
\providecommand \bibinfo  [0]{\@secondoftwo}%
\providecommand \bibfield  [0]{\@secondoftwo}%
\providecommand \translation [1]{[#1]}%
\providecommand \BibitemOpen [0]{}%
\providecommand \bibitemStop [0]{}%
\providecommand \bibitemNoStop [0]{.\EOS\space}%
\providecommand \EOS [0]{\spacefactor3000\relax}%
\providecommand \BibitemShut  [1]{\csname bibitem#1\endcsname}%
\let\auto@bib@innerbib\@empty
\bibitem [{\citenamefont {Xing}\ \emph {et~al.}(2014)\citenamefont {Xing},
  \citenamefont {Mathews}, \citenamefont {Lim}, \citenamefont {Yantara},
  \citenamefont {Liu}, \citenamefont {Sabba}, \citenamefont {Gr{\"a}tzel},
  \citenamefont {Mhaisalkar},\ and\ \citenamefont {Sum}}]{Xing14}%
  \BibitemOpen
  \bibfield  {author} {\bibinfo {author} {\bibfnamefont {G.}~\bibnamefont
  {Xing}}, \bibinfo {author} {\bibfnamefont {N.}~\bibnamefont {Mathews}},
  \bibinfo {author} {\bibfnamefont {S.~S.}\ \bibnamefont {Lim}}, \bibinfo
  {author} {\bibfnamefont {N.}~\bibnamefont {Yantara}}, \bibinfo {author}
  {\bibfnamefont {X.}~\bibnamefont {Liu}}, \bibinfo {author} {\bibfnamefont
  {D.}~\bibnamefont {Sabba}}, \bibinfo {author} {\bibfnamefont
  {M.}~\bibnamefont {Gr{\"a}tzel}}, \bibinfo {author} {\bibfnamefont
  {S.}~\bibnamefont {Mhaisalkar}},\ and\ \bibinfo {author} {\bibfnamefont
  {T.~C.}\ \bibnamefont {Sum}},\ }\bibfield  {title} {\bibinfo {title}
  {Low-temperature solution-processed wavelength-tunable perovskites for
  lasing},\ }\href {https://doi.org/10.1038/nmat3911} {\bibfield  {journal}
  {\bibinfo  {journal} {Nat Mater}\ }\textbf {\bibinfo {volume} {13}},\
  \bibinfo {pages} {476} (\bibinfo {year} {2014})},\ \bibinfo {note}
  {letter}\BibitemShut {NoStop}%
\bibitem [{\citenamefont {Green}\ \emph {et~al.}(2014)\citenamefont {Green},
  \citenamefont {Ho-Baillie},\ and\ \citenamefont {Snaith}}]{Green14}%
  \BibitemOpen
  \bibfield  {author} {\bibinfo {author} {\bibfnamefont {M.~A.}\ \bibnamefont
  {Green}}, \bibinfo {author} {\bibfnamefont {A.}~\bibnamefont {Ho-Baillie}},\
  and\ \bibinfo {author} {\bibfnamefont {H.~J.}\ \bibnamefont {Snaith}},\
  }\bibfield  {title} {\bibinfo {title} {The emergence of perovskite solar
  cells},\ }\href {https://doi.org/10.1038/nphoton.2014.134} {\bibfield
  {journal} {\bibinfo  {journal} {Nat Photon}\ }\textbf {\bibinfo {volume}
  {8}},\ \bibinfo {pages} {506} (\bibinfo {year} {2014})}\BibitemShut {NoStop}%
\bibitem [{\citenamefont {Protesescu}\ \emph {et~al.}(2015)\citenamefont
  {Protesescu}, \citenamefont {Yakunin}, \citenamefont {Bodnarchuk},
  \citenamefont {Krieg}, \citenamefont {Caputo}, \citenamefont {Hendon},
  \citenamefont {Yang}, \citenamefont {Walsh},\ and\ \citenamefont
  {Kovalenko}}]{Protesescu15}%
  \BibitemOpen
  \bibfield  {author} {\bibinfo {author} {\bibfnamefont {L.}~\bibnamefont
  {Protesescu}}, \bibinfo {author} {\bibfnamefont {S.}~\bibnamefont {Yakunin}},
  \bibinfo {author} {\bibfnamefont {M.~I.}\ \bibnamefont {Bodnarchuk}},
  \bibinfo {author} {\bibfnamefont {F.}~\bibnamefont {Krieg}}, \bibinfo
  {author} {\bibfnamefont {R.}~\bibnamefont {Caputo}}, \bibinfo {author}
  {\bibfnamefont {C.~H.}\ \bibnamefont {Hendon}}, \bibinfo {author}
  {\bibfnamefont {R.~X.}\ \bibnamefont {Yang}}, \bibinfo {author}
  {\bibfnamefont {A.}~\bibnamefont {Walsh}},\ and\ \bibinfo {author}
  {\bibfnamefont {M.~V.}\ \bibnamefont {Kovalenko}},\ }\bibfield  {title}
  {\bibinfo {title} {Nanocrystals of cesium lead halide perovskites
  ({CsPbX}\textsubscript{3}, {X} = {Cl}, {Br}, and {I}): Novel optoelectronic
  materials showing bright emission with wide color gamut},\ }\href
  {https://doi.org/10.1021/nl5048779} {\bibfield  {journal} {\bibinfo
  {journal} {Nano Letters}\ }\textbf {\bibinfo {volume} {15}},\ \bibinfo
  {pages} {3692} (\bibinfo {year} {2015})}\BibitemShut {NoStop}%
\bibitem [{\citenamefont {Yakunin}\ \emph {et~al.}(2015)\citenamefont
  {Yakunin}, \citenamefont {Protesescu}, \citenamefont {Krieg}, \citenamefont
  {Bodnarchuk}, \citenamefont {Nedelcu}, \citenamefont {Humer}, \citenamefont
  {Luca}, \citenamefont {Fiebig}, \citenamefont {Heiss},\ and\ \citenamefont
  {Kovalenko}}]{Yakunin15}%
  \BibitemOpen
  \bibfield  {author} {\bibinfo {author} {\bibfnamefont {S.}~\bibnamefont
  {Yakunin}}, \bibinfo {author} {\bibfnamefont {L.}~\bibnamefont {Protesescu}},
  \bibinfo {author} {\bibfnamefont {F.}~\bibnamefont {Krieg}}, \bibinfo
  {author} {\bibfnamefont {M.~I.}\ \bibnamefont {Bodnarchuk}}, \bibinfo
  {author} {\bibfnamefont {G.}~\bibnamefont {Nedelcu}}, \bibinfo {author}
  {\bibfnamefont {M.}~\bibnamefont {Humer}}, \bibinfo {author} {\bibfnamefont
  {G.~D.}\ \bibnamefont {Luca}}, \bibinfo {author} {\bibfnamefont
  {M.}~\bibnamefont {Fiebig}}, \bibinfo {author} {\bibfnamefont
  {W.}~\bibnamefont {Heiss}},\ and\ \bibinfo {author} {\bibfnamefont {M.~V.}\
  \bibnamefont {Kovalenko}},\ }\bibfield  {title} {\bibinfo {title}
  {Low-threshold amplified spontaneous emission and lasing from colloidal
  nanocrystals of caesium lead halide perovskites},\ }\href
  {https://www.nature.com/articles/ncomms9056} {\bibfield  {journal} {\bibinfo
  {journal} {Nature Communications}\ }\textbf {\bibinfo {volume} {6}},\
  \bibinfo {pages} {8056} (\bibinfo {year} {2015})}\BibitemShut {NoStop}%
\bibitem [{\citenamefont {Heyd}\ \emph {et~al.}(2006)\citenamefont {Heyd},
  \citenamefont {Scuseria},\ and\ \citenamefont {Ernzerhof}}]{Heyd06}%
  \BibitemOpen
  \bibfield  {author} {\bibinfo {author} {\bibfnamefont {J.}~\bibnamefont
  {Heyd}}, \bibinfo {author} {\bibfnamefont {G.~E.}\ \bibnamefont {Scuseria}},\
  and\ \bibinfo {author} {\bibfnamefont {M.}~\bibnamefont {Ernzerhof}},\
  }\bibfield  {title} {\bibinfo {title} {Erratum: “hybrid functionals based
  on a screened coulomb potential” [j. chem. phys.118, 8207 (2003)]},\ }\href
  {https://doi.org/http://dx.doi.org/10.1063/1.2204597} {\bibfield  {journal}
  {\bibinfo  {journal} {The Journal of Chemical Physics}\ }\textbf {\bibinfo
  {volume} {124}},\ \bibinfo {eid} {219906} (\bibinfo {year}
  {2006})}\BibitemShut {NoStop}%
\bibitem [{\citenamefont {Luttinger}\ and\ \citenamefont
  {Kohn}(1955)}]{Luttinger55}%
  \BibitemOpen
  \bibfield  {author} {\bibinfo {author} {\bibfnamefont {J.~M.}\ \bibnamefont
  {Luttinger}}\ and\ \bibinfo {author} {\bibfnamefont {W.}~\bibnamefont
  {Kohn}},\ }\bibfield  {title} {\bibinfo {title} {Motion of electrons and
  holes in perturbed periodic fields},\ }\href
  {https://doi.org/10.1103/PhysRev.97.869} {\bibfield  {journal} {\bibinfo
  {journal} {Phys. Rev.}\ }\textbf {\bibinfo {volume} {97}},\ \bibinfo {pages}
  {869} (\bibinfo {year} {1955})}\BibitemShut {NoStop}%
\bibitem [{\citenamefont {Zunger}(1996)}]{Zunger_PP}%
  \BibitemOpen
  \bibfield  {author} {\bibinfo {author} {\bibfnamefont {A.}~\bibnamefont
  {Zunger}},\ }\bibfield  {title} {\bibinfo {title} {First principles and
  second prniciples (semiempirical) pseudopotentials},\ }in\ \href@noop {}
  {\emph {\bibinfo {booktitle} {Quantum theory of real materials}}},\ \bibinfo
  {series} {Kluwer international series in engineering and computer science},
  Vol.\ \bibinfo {volume} {348},\ \bibinfo {editor} {edited by\ \bibinfo
  {editor} {\bibfnamefont {J.~R.}\ \bibnamefont {Chelikowsky}}\ and\ \bibinfo
  {editor} {\bibfnamefont {S.~G.}\ \bibnamefont {Louie}}}\ (\bibinfo
  {publisher} {Kluwer academic publisher},\ \bibinfo {year} {1996})\
  Chap.~\bibinfo {chapter} {13}, p.\ \bibinfo {pages} {173}\BibitemShut
  {NoStop}%
\bibitem [{\citenamefont {Benchamekh}\ \emph {et~al.}(2012)\citenamefont
  {Benchamekh}, \citenamefont {Nestoklon}, \citenamefont {Jancu},\ and\
  \citenamefont {Voisin}}]{Xavier_book}%
  \BibitemOpen
  \bibfield  {author} {\bibinfo {author} {\bibfnamefont {R.}~\bibnamefont
  {Benchamekh}}, \bibinfo {author} {\bibfnamefont {M.}~\bibnamefont
  {Nestoklon}}, \bibinfo {author} {\bibfnamefont {J.-M.}\ \bibnamefont
  {Jancu}},\ and\ \bibinfo {author} {\bibfnamefont {P.}~\bibnamefont
  {Voisin}},\ }\bibfield  {title} {\bibinfo {title} {Theory and modelling for
  the nanoscale: The $spds*$ tight binding approach},\ }in\ \href
  {https://doi.org/10.1007/978-3-642-27512-8_2} {\emph {\bibinfo {booktitle}
  {Semiconductor Modeling Techniques}}},\ \bibinfo {series} {Springer Series in
  Materials Science}, Vol.\ \bibinfo {volume} {159},\ \bibinfo {editor} {edited
  by\ \bibinfo {editor} {\bibfnamefont {X.}~\bibnamefont {Marie}}\ and\
  \bibinfo {editor} {\bibfnamefont {N.}~\bibnamefont {Balkan}}}\ (\bibinfo
  {publisher} {Springer},\ \bibinfo {year} {2012})\ Chap.~\bibinfo {chapter}
  {2}, p.~\bibinfo {pages} {19}\BibitemShut {NoStop}%
\bibitem [{\citenamefont {Even}(2015)}]{Even15}%
  \BibitemOpen
  \bibfield  {author} {\bibinfo {author} {\bibfnamefont {J.}~\bibnamefont
  {Even}},\ }\bibfield  {title} {\bibinfo {title} {Pedestrian guide to symmetry
  properties of the reference cubic structure of {3D} all-inorganic and hybrid
  perovskites},\ }\href {https://doi.org/10.1021/acs.jpclett.5b00905}
  {\bibfield  {journal} {\bibinfo  {journal} {The Journal of Physical Chemistry
  Letters}\ }\textbf {\bibinfo {volume} {6}},\ \bibinfo {pages} {2238}
  (\bibinfo {year} {2015})}\BibitemShut {NoStop}%
\bibitem [{\citenamefont {Liu}\ \emph {et~al.}(2019)\citenamefont {Liu},
  \citenamefont {Zhao}, \citenamefont {Xiao}, \citenamefont {Zhang},
  \citenamefont {Pevere}, \citenamefont {Shi}, \citenamefont {Huang},
  \citenamefont {Zhong},\ and\ \citenamefont {Sychugov}}]{Liu19}%
  \BibitemOpen
  \bibfield  {author} {\bibinfo {author} {\bibfnamefont {L.}~\bibnamefont
  {Liu}}, \bibinfo {author} {\bibfnamefont {R.}~\bibnamefont {Zhao}}, \bibinfo
  {author} {\bibfnamefont {C.}~\bibnamefont {Xiao}}, \bibinfo {author}
  {\bibfnamefont {F.}~\bibnamefont {Zhang}}, \bibinfo {author} {\bibfnamefont
  {F.}~\bibnamefont {Pevere}}, \bibinfo {author} {\bibfnamefont
  {K.}~\bibnamefont {Shi}}, \bibinfo {author} {\bibfnamefont {H.}~\bibnamefont
  {Huang}}, \bibinfo {author} {\bibfnamefont {H.}~\bibnamefont {Zhong}},\ and\
  \bibinfo {author} {\bibfnamefont {I.}~\bibnamefont {Sychugov}},\ }\bibfield
  {title} {\bibinfo {title} {Size-dependent phase transition in perovskite
  nanocrystals},\ }\href {https://doi.org/10.1021/acs.jpclett.9b02058}
  {\bibfield  {journal} {\bibinfo  {journal} {The Journal of Physical Chemistry
  Letters}\ }\textbf {\bibinfo {volume} {10}},\ \bibinfo {pages} {5451}
  (\bibinfo {year} {2019})}\BibitemShut {NoStop}%
\bibitem [{\citenamefont {Koster}\ \emph {et~al.}(1963)\citenamefont {Koster},
  \citenamefont {Dimmock}, \citenamefont {Wheeler},\ and\ \citenamefont
  {Statz}}]{bookKoster}%
  \BibitemOpen
  \bibfield  {author} {\bibinfo {author} {\bibfnamefont {G.~F.}\ \bibnamefont
  {Koster}}, \bibinfo {author} {\bibfnamefont {J.~O.}\ \bibnamefont {Dimmock}},
  \bibinfo {author} {\bibfnamefont {R.~G.}\ \bibnamefont {Wheeler}},\ and\
  \bibinfo {author} {\bibfnamefont {H.}~\bibnamefont {Statz}},\ }\href@noop {}
  {\emph {\bibinfo {title} {The Properties of the Thirty-Two Point Groups}}}\
  (\bibinfo  {publisher} {M.I.T. Press, Cambridge},\ \bibinfo {year}
  {1963})\BibitemShut {NoStop}%
\bibitem [{\citenamefont {Boyer-Richard}\ \emph {et~al.}(2016)\citenamefont
  {Boyer-Richard}, \citenamefont {Katan}, \citenamefont {Traoré},
  \citenamefont {Scholz}, \citenamefont {Jancu},\ and\ \citenamefont
  {Even}}]{Soline16}%
  \BibitemOpen
  \bibfield  {author} {\bibinfo {author} {\bibfnamefont {S.}~\bibnamefont
  {Boyer-Richard}}, \bibinfo {author} {\bibfnamefont {C.}~\bibnamefont
  {Katan}}, \bibinfo {author} {\bibfnamefont {B.}~\bibnamefont {Traoré}},
  \bibinfo {author} {\bibfnamefont {R.}~\bibnamefont {Scholz}}, \bibinfo
  {author} {\bibfnamefont {J.-M.}\ \bibnamefont {Jancu}},\ and\ \bibinfo
  {author} {\bibfnamefont {J.}~\bibnamefont {Even}},\ }\bibfield  {title}
  {\bibinfo {title} {Symmetry-based tight binding modeling of halide perovskite
  semiconductors},\ }\href {https://doi.org/10.1021/acs.jpclett.6b01749}
  {\bibfield  {journal} {\bibinfo  {journal} {The Journal of Physical Chemistry
  Letters}\ }\textbf {\bibinfo {volume} {7}},\ \bibinfo {pages} {3833}
  (\bibinfo {year} {2016})}\BibitemShut {NoStop}%
\bibitem [{\citenamefont {Jancu}\ \emph {et~al.}(1998)\citenamefont {Jancu},
  \citenamefont {Scholz}, \citenamefont {Beltram},\ and\ \citenamefont
  {Bassani}}]{Jancu98}%
  \BibitemOpen
  \bibfield  {author} {\bibinfo {author} {\bibfnamefont {J.-M.}\ \bibnamefont
  {Jancu}}, \bibinfo {author} {\bibfnamefont {R.}~\bibnamefont {Scholz}},
  \bibinfo {author} {\bibfnamefont {F.}~\bibnamefont {Beltram}},\ and\ \bibinfo
  {author} {\bibfnamefont {F.}~\bibnamefont {Bassani}},\ }\bibfield  {title}
  {\bibinfo {title} {Empirical spds* tight-binding calculation for cubic
  semiconductors: General method and material parameters},\ }\href
  {https://doi.org/10.1103/PhysRevB.57.6493} {\bibfield  {journal} {\bibinfo
  {journal} {Phys. Rev. B}\ }\textbf {\bibinfo {volume} {57}},\ \bibinfo
  {pages} {6493} (\bibinfo {year} {1998})}\BibitemShut {NoStop}%
\bibitem [{\citenamefont {Niquet}\ \emph {et~al.}(2009)\citenamefont {Niquet},
  \citenamefont {Rideau}, \citenamefont {Tavernier}, \citenamefont {Jaouen},\
  and\ \citenamefont {Blase}}]{Niquet09}%
  \BibitemOpen
  \bibfield  {author} {\bibinfo {author} {\bibfnamefont {Y.~M.}\ \bibnamefont
  {Niquet}}, \bibinfo {author} {\bibfnamefont {D.}~\bibnamefont {Rideau}},
  \bibinfo {author} {\bibfnamefont {C.}~\bibnamefont {Tavernier}}, \bibinfo
  {author} {\bibfnamefont {H.}~\bibnamefont {Jaouen}},\ and\ \bibinfo {author}
  {\bibfnamefont {X.}~\bibnamefont {Blase}},\ }\bibfield  {title} {\bibinfo
  {title} {Onsite matrix elements of the tight-binding hamiltonian of a
  strained crystal: Application to silicon, germanium, and their alloys},\
  }\href {https://doi.org/10.1103/PhysRevB.79.245201} {\bibfield  {journal}
  {\bibinfo  {journal} {Phys. Rev. B}\ }\textbf {\bibinfo {volume} {79}},\
  \bibinfo {pages} {245201} (\bibinfo {year} {2009})}\BibitemShut {NoStop}%
\bibitem [{\citenamefont {Robert}\ \emph {et~al.}(2012)\citenamefont {Robert},
  \citenamefont {Cornet}, \citenamefont {Turban}, \citenamefont {Nguyen~Thanh},
  \citenamefont {Nestoklon}, \citenamefont {Even}, \citenamefont {Jancu},
  \citenamefont {Perrin}, \citenamefont {Folliot}, \citenamefont {Rohel},
  \citenamefont {Tricot}, \citenamefont {Balocchi}, \citenamefont {Lagarde},
  \citenamefont {Marie}, \citenamefont {Bertru}, \citenamefont {Durand},\ and\
  \citenamefont {Le~Corre}}]{Robert12}%
  \BibitemOpen
  \bibfield  {author} {\bibinfo {author} {\bibfnamefont {C.}~\bibnamefont
  {Robert}}, \bibinfo {author} {\bibfnamefont {C.}~\bibnamefont {Cornet}},
  \bibinfo {author} {\bibfnamefont {P.}~\bibnamefont {Turban}}, \bibinfo
  {author} {\bibfnamefont {T.}~\bibnamefont {Nguyen~Thanh}}, \bibinfo {author}
  {\bibfnamefont {M.~O.}\ \bibnamefont {Nestoklon}}, \bibinfo {author}
  {\bibfnamefont {J.}~\bibnamefont {Even}}, \bibinfo {author} {\bibfnamefont
  {J.~M.}\ \bibnamefont {Jancu}}, \bibinfo {author} {\bibfnamefont
  {M.}~\bibnamefont {Perrin}}, \bibinfo {author} {\bibfnamefont
  {H.}~\bibnamefont {Folliot}}, \bibinfo {author} {\bibfnamefont
  {T.}~\bibnamefont {Rohel}}, \bibinfo {author} {\bibfnamefont
  {S.}~\bibnamefont {Tricot}}, \bibinfo {author} {\bibfnamefont
  {A.}~\bibnamefont {Balocchi}}, \bibinfo {author} {\bibfnamefont
  {D.}~\bibnamefont {Lagarde}}, \bibinfo {author} {\bibfnamefont
  {X.}~\bibnamefont {Marie}}, \bibinfo {author} {\bibfnamefont
  {N.}~\bibnamefont {Bertru}}, \bibinfo {author} {\bibfnamefont
  {O.}~\bibnamefont {Durand}},\ and\ \bibinfo {author} {\bibfnamefont
  {A.}~\bibnamefont {Le~Corre}},\ }\bibfield  {title} {\bibinfo {title}
  {Electronic, optical, and structural properties of {(In,Ga)As/GaP} quantum
  dots},\ }\href {https://doi.org/10.1103/PhysRevB.86.205316} {\bibfield
  {journal} {\bibinfo  {journal} {Phys. Rev. B}\ }\textbf {\bibinfo {volume}
  {86}},\ \bibinfo {pages} {205316} (\bibinfo {year} {2012})}\BibitemShut
  {NoStop}%
\bibitem [{\citenamefont {Poddubny}\ \emph {et~al.}(2012)\citenamefont
  {Poddubny}, \citenamefont {Nestoklon},\ and\ \citenamefont
  {Goupalov}}]{Poddubny12}%
  \BibitemOpen
  \bibfield  {author} {\bibinfo {author} {\bibfnamefont {A.~N.}\ \bibnamefont
  {Poddubny}}, \bibinfo {author} {\bibfnamefont {M.~O.}\ \bibnamefont
  {Nestoklon}},\ and\ \bibinfo {author} {\bibfnamefont {S.~V.}\ \bibnamefont
  {Goupalov}},\ }\bibfield  {title} {\bibinfo {title} {Anomalous suppression of
  valley splittings in lead salt nanocrystals},\ }\href
  {https://doi.org/10.1103/PhysRevB.86.035324} {\bibfield  {journal} {\bibinfo
  {journal} {Phys. Rev. B}\ }\textbf {\bibinfo {volume} {86}},\ \bibinfo
  {pages} {035324} (\bibinfo {year} {2012})}\BibitemShut {NoStop}%
\bibitem [{\citenamefont {Nestoklon}\ \emph {et~al.}(2016)\citenamefont
  {Nestoklon}, \citenamefont {Poddubny}, \citenamefont {Voisin},\ and\
  \citenamefont {Dohnalova}}]{Nestoklon16_Ge}%
  \BibitemOpen
  \bibfield  {author} {\bibinfo {author} {\bibfnamefont {M.~O.}\ \bibnamefont
  {Nestoklon}}, \bibinfo {author} {\bibfnamefont {A.~N.}\ \bibnamefont
  {Poddubny}}, \bibinfo {author} {\bibfnamefont {P.}~\bibnamefont {Voisin}},\
  and\ \bibinfo {author} {\bibfnamefont {K.}~\bibnamefont {Dohnalova}},\
  }\bibfield  {title} {\bibinfo {title} {Tuning optical properties of {Ge}
  nanocrystals by {Si} shell},\ }\href
  {https://doi.org/10.1021/acs.jpcc.6b05753} {\bibfield  {journal} {\bibinfo
  {journal} {The Journal of Physical Chemistry C}\ }\textbf {\bibinfo {volume}
  {120}},\ \bibinfo {pages} {18901} (\bibinfo {year} {2016})}\BibitemShut
  {NoStop}%
\bibitem [{\citenamefont {Robert}\ \emph {et~al.}(2016)\citenamefont {Robert},
  \citenamefont {Pereira Da~Silva}, \citenamefont {Nestoklon}, \citenamefont
  {Alonso}, \citenamefont {Turban}, \citenamefont {Jancu}, \citenamefont
  {Even}, \citenamefont {Carr{\`e}re}, \citenamefont {Balocchi}, \citenamefont
  {Koenraad}, \citenamefont {Marie}, \citenamefont {Durand}, \citenamefont
  {Go{\~n}i},\ and\ \citenamefont {Cornet}}]{Robert16}%
  \BibitemOpen
  \bibfield  {author} {\bibinfo {author} {\bibfnamefont {C.}~\bibnamefont
  {Robert}}, \bibinfo {author} {\bibfnamefont {K.}~\bibnamefont {Pereira
  Da~Silva}}, \bibinfo {author} {\bibfnamefont {M.~O.}\ \bibnamefont
  {Nestoklon}}, \bibinfo {author} {\bibfnamefont {M.~I.}\ \bibnamefont
  {Alonso}}, \bibinfo {author} {\bibfnamefont {P.}~\bibnamefont {Turban}},
  \bibinfo {author} {\bibfnamefont {J.-M.}\ \bibnamefont {Jancu}}, \bibinfo
  {author} {\bibfnamefont {J.}~\bibnamefont {Even}}, \bibinfo {author}
  {\bibfnamefont {H.}~\bibnamefont {Carr{\`e}re}}, \bibinfo {author}
  {\bibfnamefont {A.}~\bibnamefont {Balocchi}}, \bibinfo {author}
  {\bibfnamefont {P.~M.}\ \bibnamefont {Koenraad}}, \bibinfo {author}
  {\bibfnamefont {X.}~\bibnamefont {Marie}}, \bibinfo {author} {\bibfnamefont
  {O.}~\bibnamefont {Durand}}, \bibinfo {author} {\bibfnamefont {A.~R.}\
  \bibnamefont {Go{\~n}i}},\ and\ \bibinfo {author} {\bibfnamefont
  {C.}~\bibnamefont {Cornet}},\ }\bibfield  {title} {\bibinfo {title}
  {Electronic wave functions and optical transitions in {(In,Ga)As/GaP} quantum
  dots},\ }\href {https://doi.org/10.1103/PhysRevB.94.075445} {\bibfield
  {journal} {\bibinfo  {journal} {Phys. Rev. B}\ }\textbf {\bibinfo {volume}
  {94}},\ \bibinfo {pages} {075445} (\bibinfo {year} {2016})}\BibitemShut
  {NoStop}%
\bibitem [{\citenamefont {Benchamekh}\ \emph {et~al.}(2016)\citenamefont
  {Benchamekh}, \citenamefont {Schulz},\ and\ \citenamefont
  {O'Reilly}}]{Benchamekh16}%
  \BibitemOpen
  \bibfield  {author} {\bibinfo {author} {\bibfnamefont {R.}~\bibnamefont
  {Benchamekh}}, \bibinfo {author} {\bibfnamefont {S.}~\bibnamefont {Schulz}},\
  and\ \bibinfo {author} {\bibfnamefont {E.~P.}\ \bibnamefont {O'Reilly}},\
  }\bibfield  {title} {\bibinfo {title} {Theoretical analysis of influence of
  random alloy fluctuations on the optoelectronic properties of site-controlled
  (111)-oriented {InGaAs/GaAs} quantum dots},\ }\href
  {https://doi.org/10.1103/PhysRevB.94.125308} {\bibfield  {journal} {\bibinfo
  {journal} {Phys. Rev. B}\ }\textbf {\bibinfo {volume} {94}},\ \bibinfo
  {pages} {125308} (\bibinfo {year} {2016})}\BibitemShut {NoStop}%
\bibitem [{\citenamefont {Tan}\ \emph {et~al.}(2016)\citenamefont {Tan},
  \citenamefont {Povolotskyi}, \citenamefont {Kubis}, \citenamefont {Boykin},\
  and\ \citenamefont {Klimeck}}]{Tan16}%
  \BibitemOpen
  \bibfield  {author} {\bibinfo {author} {\bibfnamefont {Y.}~\bibnamefont
  {Tan}}, \bibinfo {author} {\bibfnamefont {M.}~\bibnamefont {Povolotskyi}},
  \bibinfo {author} {\bibfnamefont {T.}~\bibnamefont {Kubis}}, \bibinfo
  {author} {\bibfnamefont {T.~B.}\ \bibnamefont {Boykin}},\ and\ \bibinfo
  {author} {\bibfnamefont {G.}~\bibnamefont {Klimeck}},\ }\bibfield  {title}
  {\bibinfo {title} {Transferable tight-binding model for strained group iv and
  iii-v materials and heterostructures},\ }\href
  {https://doi.org/10.1103/PhysRevB.94.045311} {\bibfield  {journal} {\bibinfo
  {journal} {Phys. Rev. B}\ }\textbf {\bibinfo {volume} {94}},\ \bibinfo
  {pages} {045311} (\bibinfo {year} {2016})}\BibitemShut {NoStop}%
\bibitem [{\citenamefont {Tan}\ \emph {et~al.}(2015)\citenamefont {Tan},
  \citenamefont {Povolotskyi}, \citenamefont {Kubis}, \citenamefont {Boykin},\
  and\ \citenamefont {Klimeck}}]{Tan15}%
  \BibitemOpen
  \bibfield  {author} {\bibinfo {author} {\bibfnamefont {Y.~P.}\ \bibnamefont
  {Tan}}, \bibinfo {author} {\bibfnamefont {M.}~\bibnamefont {Povolotskyi}},
  \bibinfo {author} {\bibfnamefont {T.}~\bibnamefont {Kubis}}, \bibinfo
  {author} {\bibfnamefont {T.~B.}\ \bibnamefont {Boykin}},\ and\ \bibinfo
  {author} {\bibfnamefont {G.}~\bibnamefont {Klimeck}},\ }\bibfield  {title}
  {\bibinfo {title} {Tight-binding analysis of {Si} and {GaAs} ultrathin bodies
  with subatomic wave-function resolution},\ }\href
  {https://doi.org/10.1103/PhysRevB.92.085301} {\bibfield  {journal} {\bibinfo
  {journal} {Phys. Rev. B}\ }\textbf {\bibinfo {volume} {92}},\ \bibinfo
  {pages} {085301} (\bibinfo {year} {2015})}\BibitemShut {NoStop}%
\bibitem [{\citenamefont {Gresch}\ \emph {et~al.}(2018)\citenamefont {Gresch},
  \citenamefont {Wu}, \citenamefont {Winkler}, \citenamefont {H\"auselmann},
  \citenamefont {Troyer},\ and\ \citenamefont {Soluyanov}}]{Gresch18}%
  \BibitemOpen
  \bibfield  {author} {\bibinfo {author} {\bibfnamefont {D.}~\bibnamefont
  {Gresch}}, \bibinfo {author} {\bibfnamefont {Q.}~\bibnamefont {Wu}}, \bibinfo
  {author} {\bibfnamefont {G.~W.}\ \bibnamefont {Winkler}}, \bibinfo {author}
  {\bibfnamefont {R.}~\bibnamefont {H\"auselmann}}, \bibinfo {author}
  {\bibfnamefont {M.}~\bibnamefont {Troyer}},\ and\ \bibinfo {author}
  {\bibfnamefont {A.~A.}\ \bibnamefont {Soluyanov}},\ }\bibfield  {title}
  {\bibinfo {title} {Automated construction of symmetrized wannier-like
  tight-binding models from ab initio calculations},\ }\href
  {https://doi.org/10.1103/PhysRevMaterials.2.103805} {\bibfield  {journal}
  {\bibinfo  {journal} {Phys. Rev. Materials}\ }\textbf {\bibinfo {volume}
  {2}},\ \bibinfo {pages} {103805} (\bibinfo {year} {2018})}\BibitemShut
  {NoStop}%
\bibitem [{\citenamefont {Lihm}\ and\ \citenamefont {Park}(2019)}]{Lihm19}%
  \BibitemOpen
  \bibfield  {author} {\bibinfo {author} {\bibfnamefont {J.-M.}\ \bibnamefont
  {Lihm}}\ and\ \bibinfo {author} {\bibfnamefont {C.-H.}\ \bibnamefont
  {Park}},\ }\bibfield  {title} {\bibinfo {title} {Reliable methods for
  seamless stitching of tight-binding models based on maximally localized
  wannier functions},\ }\href {https://doi.org/10.1103/PhysRevB.99.125117}
  {\bibfield  {journal} {\bibinfo  {journal} {Phys. Rev. B}\ }\textbf {\bibinfo
  {volume} {99}},\ \bibinfo {pages} {125117} (\bibinfo {year}
  {2019})}\BibitemShut {NoStop}%
\bibitem [{\citenamefont {{Di Vito}}\ \emph {et~al.}(2020)\citenamefont {{Di
  Vito}}, \citenamefont {{Pecchia}}, \citenamefont {{Auf der Maur}},\ and\
  \citenamefont {{Di Carlo}}}]{DiVito20}%
  \BibitemOpen
  \bibfield  {author} {\bibinfo {author} {\bibfnamefont {A.}~\bibnamefont {{Di
  Vito}}}, \bibinfo {author} {\bibfnamefont {A.}~\bibnamefont {{Pecchia}}},
  \bibinfo {author} {\bibfnamefont {M.}~\bibnamefont {{Auf der Maur}}},\ and\
  \bibinfo {author} {\bibfnamefont {A.}~\bibnamefont {{Di Carlo}}},\ }\bibfield
   {title} {\bibinfo {title} {Tight binding parameterization through particle
  swarm optimization algorithm},\ }in\ \href
  {https://doi.org/10.1109/NUSOD49422.2020.9217665} {\emph {\bibinfo
  {booktitle} {2020 International Conference on Numerical Simulation of
  Optoelectronic Devices (NUSOD)}}}\ (\bibinfo {year} {2020})\ pp.\ \bibinfo
  {pages} {113--114}\BibitemShut {NoStop}%
\bibitem [{\citenamefont {Tran}\ and\ \citenamefont {Blaha}(2009)}]{Tran09}%
  \BibitemOpen
  \bibfield  {author} {\bibinfo {author} {\bibfnamefont {F.}~\bibnamefont
  {Tran}}\ and\ \bibinfo {author} {\bibfnamefont {P.}~\bibnamefont {Blaha}},\
  }\bibfield  {title} {\bibinfo {title} {Accurate band gaps of semiconductors
  and insulators with a semilocal exchange-correlation potential},\ }\href
  {https://doi.org/10.1103/PhysRevLett.102.226401} {\bibfield  {journal}
  {\bibinfo  {journal} {Phys. Rev. Lett.}\ }\textbf {\bibinfo {volume} {102}},\
  \bibinfo {pages} {226401} (\bibinfo {year} {2009})}\BibitemShut {NoStop}%
\bibitem [{\citenamefont {Jishi}\ \emph {et~al.}(2014)\citenamefont {Jishi},
  \citenamefont {Ta},\ and\ \citenamefont {Sharif}}]{Jishi14}%
  \BibitemOpen
  \bibfield  {author} {\bibinfo {author} {\bibfnamefont {R.~A.}\ \bibnamefont
  {Jishi}}, \bibinfo {author} {\bibfnamefont {O.~B.}\ \bibnamefont {Ta}},\ and\
  \bibinfo {author} {\bibfnamefont {A.~A.}\ \bibnamefont {Sharif}},\ }\bibfield
   {title} {\bibinfo {title} {Modeling of lead halide perovskites for
  photovoltaic applications},\ }\href {https://doi.org/10.1021/jp5050145}
  {\bibfield  {journal} {\bibinfo  {journal} {The Journal of Physical Chemistry
  C}\ }\textbf {\bibinfo {volume} {118}},\ \bibinfo {pages} {28344} (\bibinfo
  {year} {2014})}\BibitemShut {NoStop}%
\bibitem [{\citenamefont {Camargo-Mart{\'{\i}}nez}\ and\ \citenamefont
  {Baquero}(2012)}]{Camargo12}%
  \BibitemOpen
  \bibfield  {author} {\bibinfo {author} {\bibfnamefont {J.~A.}\ \bibnamefont
  {Camargo-Mart{\'{\i}}nez}}\ and\ \bibinfo {author} {\bibfnamefont
  {R.}~\bibnamefont {Baquero}},\ }\bibfield  {title} {\bibinfo {title}
  {Performance of the modified {Becke-Johnson} potential for semiconductors},\
  }\href {https://doi.org/10.1103/PhysRevB.86.195106} {\bibfield  {journal}
  {\bibinfo  {journal} {Phys. Rev. B}\ }\textbf {\bibinfo {volume} {86}},\
  \bibinfo {pages} {195106} (\bibinfo {year} {2012})}\BibitemShut {NoStop}%
\bibitem [{\citenamefont {Wang}\ \emph {et~al.}(2013)\citenamefont {Wang},
  \citenamefont {Yin}, \citenamefont {Cao}, \citenamefont {Zahid},
  \citenamefont {Zhu}, \citenamefont {Liu}, \citenamefont {Wang},\ and\
  \citenamefont {Guo}}]{Wang13}%
  \BibitemOpen
  \bibfield  {author} {\bibinfo {author} {\bibfnamefont {Y.}~\bibnamefont
  {Wang}}, \bibinfo {author} {\bibfnamefont {H.}~\bibnamefont {Yin}}, \bibinfo
  {author} {\bibfnamefont {R.}~\bibnamefont {Cao}}, \bibinfo {author}
  {\bibfnamefont {F.}~\bibnamefont {Zahid}}, \bibinfo {author} {\bibfnamefont
  {Y.}~\bibnamefont {Zhu}}, \bibinfo {author} {\bibfnamefont {L.}~\bibnamefont
  {Liu}}, \bibinfo {author} {\bibfnamefont {J.}~\bibnamefont {Wang}},\ and\
  \bibinfo {author} {\bibfnamefont {H.}~\bibnamefont {Guo}},\ }\bibfield
  {title} {\bibinfo {title} {Electronic structure of {III-V} zinc-blende
  semiconductors from first principles},\ }\href
  {https://doi.org/10.1103/PhysRevB.87.235203} {\bibfield  {journal} {\bibinfo
  {journal} {Phys. Rev. B}\ }\textbf {\bibinfo {volume} {87}},\ \bibinfo
  {pages} {235203} (\bibinfo {year} {2013})}\BibitemShut {NoStop}%
\bibitem [{\citenamefont {Cardona}(2001)}]{Cardona01}%
  \BibitemOpen
  \bibfield  {author} {\bibinfo {author} {\bibfnamefont {M.}~\bibnamefont
  {Cardona}},\ }\bibfield  {title} {\bibinfo {title} {Renormalization of the
  optical response of semiconductors by electron–phonon interaction},\ }\href
  {https://doi.org/https://doi.org/10.1002/1521-396X(200112)188:4<1209::AID-PSSA1209>3.0.CO;2-2}
  {\bibfield  {journal} {\bibinfo  {journal} {physica status solidi (a)}\
  }\textbf {\bibinfo {volume} {188}},\ \bibinfo {pages} {1209} (\bibinfo {year}
  {2001})}\BibitemShut {NoStop}%
\bibitem [{\citenamefont {Wiktor}\ \emph {et~al.}(2017)\citenamefont {Wiktor},
  \citenamefont {Rothlisberger},\ and\ \citenamefont {Pasquarello}}]{Wiktor17}%
  \BibitemOpen
  \bibfield  {author} {\bibinfo {author} {\bibfnamefont {J.}~\bibnamefont
  {Wiktor}}, \bibinfo {author} {\bibfnamefont {U.}~\bibnamefont
  {Rothlisberger}},\ and\ \bibinfo {author} {\bibfnamefont {A.}~\bibnamefont
  {Pasquarello}},\ }\bibfield  {title} {\bibinfo {title} {Predictive
  determination of band gaps of inorganic halide perovskites},\ }\href
  {https://doi.org/10.1021/acs.jpclett.7b02648} {\bibfield  {journal} {\bibinfo
   {journal} {The Journal of Physical Chemistry Letters}\ }\textbf {\bibinfo
  {volume} {8}},\ \bibinfo {pages} {5507} (\bibinfo {year} {2017})}\BibitemShut
  {NoStop}%
\bibitem [{\citenamefont {Patrick}\ \emph {et~al.}(2015)\citenamefont
  {Patrick}, \citenamefont {Jacobsen},\ and\ \citenamefont
  {Thygesen}}]{Patrick15}%
  \BibitemOpen
  \bibfield  {author} {\bibinfo {author} {\bibfnamefont {C.~E.}\ \bibnamefont
  {Patrick}}, \bibinfo {author} {\bibfnamefont {K.~W.}\ \bibnamefont
  {Jacobsen}},\ and\ \bibinfo {author} {\bibfnamefont {K.~S.}\ \bibnamefont
  {Thygesen}},\ }\bibfield  {title} {\bibinfo {title} {Anharmonic stabilization
  and band gap renormalization in the perovskite {CsSnI}\textsubscript{3}},\
  }\href {https://doi.org/10.1103/PhysRevB.92.201205} {\bibfield  {journal}
  {\bibinfo  {journal} {Phys. Rev. B}\ }\textbf {\bibinfo {volume} {92}},\
  \bibinfo {pages} {201205} (\bibinfo {year} {2015})}\BibitemShut {NoStop}%
\bibitem [{\citenamefont {Zacharias}\ \emph {et~al.}(2020)\citenamefont
  {Zacharias}, \citenamefont {Scheffler},\ and\ \citenamefont
  {Carbogno}}]{Zacharias20}%
  \BibitemOpen
  \bibfield  {author} {\bibinfo {author} {\bibfnamefont {M.}~\bibnamefont
  {Zacharias}}, \bibinfo {author} {\bibfnamefont {M.}~\bibnamefont
  {Scheffler}},\ and\ \bibinfo {author} {\bibfnamefont {C.}~\bibnamefont
  {Carbogno}},\ }\bibfield  {title} {\bibinfo {title} {Fully anharmonic
  nonperturbative theory of vibronically renormalized electronic band
  structures},\ }\href {https://doi.org/10.1103/PhysRevB.102.045126} {\bibfield
   {journal} {\bibinfo  {journal} {Phys. Rev. B}\ }\textbf {\bibinfo {volume}
  {102}},\ \bibinfo {pages} {045126} (\bibinfo {year} {2020})}\BibitemShut
  {NoStop}%
\bibitem [{\citenamefont {Blaha}\ \emph {et~al.}(2020)\citenamefont {Blaha},
  \citenamefont {Schwarz}, \citenamefont {Tran}, \citenamefont {Laskowski},
  \citenamefont {Madsen},\ and\ \citenamefont {Marks}}]{wien2k}%
  \BibitemOpen
  \bibfield  {author} {\bibinfo {author} {\bibfnamefont {P.}~\bibnamefont
  {Blaha}}, \bibinfo {author} {\bibfnamefont {K.}~\bibnamefont {Schwarz}},
  \bibinfo {author} {\bibfnamefont {F.}~\bibnamefont {Tran}}, \bibinfo {author}
  {\bibfnamefont {R.}~\bibnamefont {Laskowski}}, \bibinfo {author}
  {\bibfnamefont {G.~K.~H.}\ \bibnamefont {Madsen}},\ and\ \bibinfo {author}
  {\bibfnamefont {L.~D.}\ \bibnamefont {Marks}},\ }\bibfield  {title} {\bibinfo
  {title} {{WIEN2k}: An {APW+lo} program for calculating the properties of
  solids},\ }\href {https://doi.org/10.1063/1.5143061} {\bibfield  {journal}
  {\bibinfo  {journal} {J. Chem. Phys.}\ }\textbf {\bibinfo {volume} {152}},\
  \bibinfo {pages} {074101} (\bibinfo {year} {2020})}\BibitemShut {NoStop}%
\bibitem [{\citenamefont {Kune{\v{s}}}\ \emph {et~al.}(2001)\citenamefont
  {Kune{\v{s}}}, \citenamefont {Nov{\'a}k}, \citenamefont {Schmid},
  \citenamefont {Blaha},\ and\ \citenamefont {Schwarz}}]{Kunes01}%
  \BibitemOpen
  \bibfield  {author} {\bibinfo {author} {\bibfnamefont {J.}~\bibnamefont
  {Kune{\v{s}}}}, \bibinfo {author} {\bibfnamefont {P.}~\bibnamefont
  {Nov{\'a}k}}, \bibinfo {author} {\bibfnamefont {R.}~\bibnamefont {Schmid}},
  \bibinfo {author} {\bibfnamefont {P.}~\bibnamefont {Blaha}},\ and\ \bibinfo
  {author} {\bibfnamefont {K.}~\bibnamefont {Schwarz}},\ }\bibfield  {title}
  {\bibinfo {title} {Electronic structure of fcc {Th}: Spin-orbit calculation
  with ${6p}_{1/2}$ local orbital extension},\ }\href
  {https://doi.org/10.1103/PhysRevB.64.153102} {\bibfield  {journal} {\bibinfo
  {journal} {Phys. Rev. B}\ }\textbf {\bibinfo {volume} {64}},\ \bibinfo
  {pages} {153102} (\bibinfo {year} {2001})}\BibitemShut {NoStop}%
\bibitem [{\citenamefont {Yuan}\ \emph {et~al.}(2020)\citenamefont {Yuan},
  \citenamefont {Yuan}, \citenamefont {Hu}, \citenamefont {Yu}, \citenamefont
  {Li}, \citenamefont {Barea}, \citenamefont {Bisquert},\ and\ \citenamefont
  {Meng}}]{Yuan20}%
  \BibitemOpen
  \bibfield  {author} {\bibinfo {author} {\bibfnamefont {M.}~\bibnamefont
  {Yuan}}, \bibinfo {author} {\bibfnamefont {L.}~\bibnamefont {Yuan}}, \bibinfo
  {author} {\bibfnamefont {Z.}~\bibnamefont {Hu}}, \bibinfo {author}
  {\bibfnamefont {Z.}~\bibnamefont {Yu}}, \bibinfo {author} {\bibfnamefont
  {H.}~\bibnamefont {Li}}, \bibinfo {author} {\bibfnamefont {E.~M.}\
  \bibnamefont {Barea}}, \bibinfo {author} {\bibfnamefont {J.}~\bibnamefont
  {Bisquert}},\ and\ \bibinfo {author} {\bibfnamefont {X.}~\bibnamefont
  {Meng}},\ }\bibfield  {title} {\bibinfo {title} {In situ spectroscopic
  ellipsometry for thermochromic {CsPbI}\textsubscript{3} phase evolution
  portfolio},\ }\href {https://doi.org/10.1021/acs.jpcc.0c01231} {\bibfield
  {journal} {\bibinfo  {journal} {The Journal of Physical Chemistry C}\
  }\textbf {\bibinfo {volume} {124}},\ \bibinfo {pages} {8008} (\bibinfo {year}
  {2020})}\BibitemShut {NoStop}%
\bibitem [{\citenamefont {Slater}\ and\ \citenamefont
  {Koster}(1954)}]{Slater54}%
  \BibitemOpen
  \bibfield  {author} {\bibinfo {author} {\bibfnamefont {J.~C.}\ \bibnamefont
  {Slater}}\ and\ \bibinfo {author} {\bibfnamefont {G.~F.}\ \bibnamefont
  {Koster}},\ }\bibfield  {title} {\bibinfo {title} {Simplified lcao method for
  the periodic potential problem},\ }\href
  {https://doi.org/10.1103/PhysRev.94.1498} {\bibfield  {journal} {\bibinfo
  {journal} {Phys. Rev.}\ }\textbf {\bibinfo {volume} {94}},\ \bibinfo {pages}
  {1498} (\bibinfo {year} {1954})}\BibitemShut {NoStop}%
\bibitem [{\citenamefont {Vogl}\ \emph {et~al.}(1983)\citenamefont {Vogl},
  \citenamefont {Hjalmarson},\ and\ \citenamefont {Dow}}]{Vogl83}%
  \BibitemOpen
  \bibfield  {author} {\bibinfo {author} {\bibfnamefont {P.}~\bibnamefont
  {Vogl}}, \bibinfo {author} {\bibfnamefont {H.~P.}\ \bibnamefont
  {Hjalmarson}},\ and\ \bibinfo {author} {\bibfnamefont {J.~D.}\ \bibnamefont
  {Dow}},\ }\bibfield  {title} {\bibinfo {title} {A semi-empirical
  tight-binding theory of the electronic structure of semiconductors†},\
  }\href {https://doi.org/https://doi.org/10.1016/0022-3697(83)90064-1}
  {\bibfield  {journal} {\bibinfo  {journal} {Journal of Physics and Chemistry
  of Solids}\ }\textbf {\bibinfo {volume} {44}},\ \bibinfo {pages} {365 }
  (\bibinfo {year} {1983})}\BibitemShut {NoStop}%
\bibitem [{\citenamefont {Belolipetsky}\ \emph {et~al.}(2018)\citenamefont
  {Belolipetsky}, \citenamefont {Nestoklon},\ and\ \citenamefont
  {Yassievich}}]{Belolipetsky18}%
  \BibitemOpen
  \bibfield  {author} {\bibinfo {author} {\bibfnamefont {A.~V.}\ \bibnamefont
  {Belolipetsky}}, \bibinfo {author} {\bibfnamefont {M.~O.}\ \bibnamefont
  {Nestoklon}},\ and\ \bibinfo {author} {\bibfnamefont {I.~N.}\ \bibnamefont
  {Yassievich}},\ }\bibfield  {title} {\bibinfo {title} {Simulation of electron
  and hole states in {Si} nanocrystals in a {SiO}\textsubscript{2} matrix:
  Choice of parameters of the empirical tight-binding method},\ }\href
  {https://doi.org/10.1134/S1063782618100020} {\bibfield  {journal} {\bibinfo
  {journal} {Semiconductors}\ }\textbf {\bibinfo {volume} {52}},\ \bibinfo
  {pages} {1264} (\bibinfo {year} {2018})},\ \bibinfo {note}
  {[\href{https://doi.org/10.21883/FTP.2018.10.46454.8859}{Fizika i Tekhnika
  Poluprovodnikov {\bf{52}}, 1145 (2018)}]}\BibitemShut {NoStop}%
\bibitem [{\citenamefont {Boykin}\ \emph {et~al.}(2019)\citenamefont {Boykin},
  \citenamefont {Sarangapani},\ and\ \citenamefont {Klimeck}}]{Boykin19}%
  \BibitemOpen
  \bibfield  {author} {\bibinfo {author} {\bibfnamefont {T.~B.}\ \bibnamefont
  {Boykin}}, \bibinfo {author} {\bibfnamefont {P.}~\bibnamefont
  {Sarangapani}},\ and\ \bibinfo {author} {\bibfnamefont {G.}~\bibnamefont
  {Klimeck}},\ }\bibfield  {title} {\bibinfo {title} {Non-orthogonal
  tight-binding models: Problems and possible remedies for realistic nano-scale
  devices},\ }\href {https://doi.org/10.1063/1.5056178} {\bibfield  {journal}
  {\bibinfo  {journal} {Journal of Applied Physics}\ }\textbf {\bibinfo
  {volume} {125}},\ \bibinfo {pages} {144302} (\bibinfo {year}
  {2019})}\BibitemShut {NoStop}%
\bibitem [{\citenamefont {Yang}\ \emph
  {et~al.}(2019{\natexlab{a}})\citenamefont {Yang}, \citenamefont {Cao},
  \citenamefont {Zhong}, \citenamefont {Li}, \citenamefont {Zhang},\ and\
  \citenamefont {Zhang}}]{Yang19}%
  \BibitemOpen
  \bibfield  {author} {\bibinfo {author} {\bibfnamefont {D.}~\bibnamefont
  {Yang}}, \bibinfo {author} {\bibfnamefont {M.}~\bibnamefont {Cao}}, \bibinfo
  {author} {\bibfnamefont {Q.}~\bibnamefont {Zhong}}, \bibinfo {author}
  {\bibfnamefont {P.}~\bibnamefont {Li}}, \bibinfo {author} {\bibfnamefont
  {X.}~\bibnamefont {Zhang}},\ and\ \bibinfo {author} {\bibfnamefont
  {Q.}~\bibnamefont {Zhang}},\ }\bibfield  {title} {\bibinfo {title}
  {All-inorganic cesium lead halide perovskite nanocrystals: synthesis{,}
  surface engineering and applications},\ }\href
  {https://doi.org/10.1039/C8TC04381G} {\bibfield  {journal} {\bibinfo
  {journal} {J. Mater. Chem. C}\ }\textbf {\bibinfo {volume} {7}},\ \bibinfo
  {pages} {757} (\bibinfo {year} {2019}{\natexlab{a}})}\BibitemShut {NoStop}%
\bibitem [{\citenamefont {Momma}\ and\ \citenamefont
  {Izumi}(2011)}]{momma2011vesta}%
  \BibitemOpen
  \bibfield  {author} {\bibinfo {author} {\bibfnamefont {K.}~\bibnamefont
  {Momma}}\ and\ \bibinfo {author} {\bibfnamefont {F.}~\bibnamefont {Izumi}},\
  }\bibfield  {title} {\bibinfo {title} {Vesta 3 for three-dimensional
  visualization of crystal, volumetric and morphology data},\ }\href@noop {}
  {\bibfield  {journal} {\bibinfo  {journal} {Journal of applied
  crystallography}\ }\textbf {\bibinfo {volume} {44}},\ \bibinfo {pages} {1272}
  (\bibinfo {year} {2011})}\BibitemShut {NoStop}%
\bibitem [{\citenamefont {Katan}\ \emph {et~al.}(2019)\citenamefont {Katan},
  \citenamefont {Mercier},\ and\ \citenamefont {Even}}]{Katan19}%
  \BibitemOpen
  \bibfield  {author} {\bibinfo {author} {\bibfnamefont {C.}~\bibnamefont
  {Katan}}, \bibinfo {author} {\bibfnamefont {N.}~\bibnamefont {Mercier}},\
  and\ \bibinfo {author} {\bibfnamefont {J.}~\bibnamefont {Even}},\ }\bibfield
  {title} {\bibinfo {title} {Quantum and dielectric confinement effects in
  lower-dimensional hybrid perovskite semiconductors},\ }\href
  {https://doi.org/10.1021/acs.chemrev.8b00417} {\bibfield  {journal} {\bibinfo
   {journal} {Chemical Reviews}\ }\textbf {\bibinfo {volume} {119}},\ \bibinfo
  {pages} {3140} (\bibinfo {year} {2019})}\BibitemShut {NoStop}%
\bibitem [{\citenamefont {Benchamekh}\ \emph {et~al.}(2014)\citenamefont
  {Benchamekh}, \citenamefont {Gippius}, \citenamefont {Even}, \citenamefont
  {Nestoklon}, \citenamefont {Jancu}, \citenamefont {Ithurria}, \citenamefont
  {Dubertret}, \citenamefont {Efros},\ and\ \citenamefont
  {Voisin}}]{Benchamekh14}%
  \BibitemOpen
  \bibfield  {author} {\bibinfo {author} {\bibfnamefont {R.}~\bibnamefont
  {Benchamekh}}, \bibinfo {author} {\bibfnamefont {N.~A.}\ \bibnamefont
  {Gippius}}, \bibinfo {author} {\bibfnamefont {J.}~\bibnamefont {Even}},
  \bibinfo {author} {\bibfnamefont {M.~O.}\ \bibnamefont {Nestoklon}}, \bibinfo
  {author} {\bibfnamefont {J.-M.}\ \bibnamefont {Jancu}}, \bibinfo {author}
  {\bibfnamefont {S.}~\bibnamefont {Ithurria}}, \bibinfo {author}
  {\bibfnamefont {B.}~\bibnamefont {Dubertret}}, \bibinfo {author}
  {\bibfnamefont {A.~L.}\ \bibnamefont {Efros}},\ and\ \bibinfo {author}
  {\bibfnamefont {P.}~\bibnamefont {Voisin}},\ }\bibfield  {title} {\bibinfo
  {title} {Tight-binding calculations of image-charge effects in colloidal
  nanoscale platelets of {CdSe}},\ }\href
  {https://doi.org/10.1103/PhysRevB.89.035307} {\bibfield  {journal} {\bibinfo
  {journal} {Phys. Rev. B}\ }\textbf {\bibinfo {volume} {89}},\ \bibinfo
  {pages} {035307} (\bibinfo {year} {2014})}\BibitemShut {NoStop}%
\bibitem [{\citenamefont {Cho}\ and\ \citenamefont {Berkelbach}(2019)}]{Cho19}%
  \BibitemOpen
  \bibfield  {author} {\bibinfo {author} {\bibfnamefont {Y.}~\bibnamefont
  {Cho}}\ and\ \bibinfo {author} {\bibfnamefont {T.~C.}\ \bibnamefont
  {Berkelbach}},\ }\bibfield  {title} {\bibinfo {title} {Optical properties of
  layered hybrid organic–inorganic halide perovskites: A tight-binding
  {GW-BSE} study},\ }\href {https://doi.org/10.1021/acs.jpclett.9b02491}
  {\bibfield  {journal} {\bibinfo  {journal} {The Journal of Physical Chemistry
  Letters}\ }\textbf {\bibinfo {volume} {10}},\ \bibinfo {pages} {6189}
  (\bibinfo {year} {2019})}\BibitemShut {NoStop}%
\bibitem [{\citenamefont {Yang}\ \emph
  {et~al.}(2019{\natexlab{b}})\citenamefont {Yang}, \citenamefont {Wang},
  \citenamefont {Pan}, \citenamefont {Zhou}, \citenamefont {Kong},\ and\
  \citenamefont {Ji}}]{YangWang19}%
  \BibitemOpen
  \bibfield  {author} {\bibinfo {author} {\bibfnamefont {F.}~\bibnamefont
  {Yang}}, \bibinfo {author} {\bibfnamefont {C.}~\bibnamefont {Wang}}, \bibinfo
  {author} {\bibfnamefont {Y.}~\bibnamefont {Pan}}, \bibinfo {author}
  {\bibfnamefont {X.}~\bibnamefont {Zhou}}, \bibinfo {author} {\bibfnamefont
  {X.}~\bibnamefont {Kong}},\ and\ \bibinfo {author} {\bibfnamefont
  {W.}~\bibnamefont {Ji}},\ }\bibfield  {title} {\bibinfo {title} {Surface
  stabilized cubic phase of {CsPbI}\textsubscript{3} and
  {CsPbBr}\textsubscript{3} at room temperature},\ }\href
  {https://doi.org/10.1088/1674-1056/28/5/056402} {\bibfield  {journal}
  {\bibinfo  {journal} {Chinese Physics B}\ }\textbf {\bibinfo {volume} {28}},\
  \bibinfo {pages} {056402} (\bibinfo {year} {2019}{\natexlab{b}})}\BibitemShut
  {NoStop}%
\end{thebibliography}%

\end{document}